\documentclass[%        Class options:
aps,%                   American Physical Society
prd,%                   Physical Review D
tightenlines,%          Single spaced lines
superscriptaddress,%   	Authors' addresses linked with superscripts
nofootinbib,%           Does not treat footnotes as references
floatfix]%              Fixes float errors
{revtex4}%            REVTEX 4 Package used
\usepackage{graphicx}%  Default Latex 2eps package for embedding figures (prefers eps figures)
\usepackage{subfigure}
\usepackage{bm}
\begin{document}

\title{		Mantle geoneutrinos in KamLAND and Borexino}

\author{ 			G.\ Fiorentini}
\affiliation{Dipartimento di Fisica, Universit\`a di Ferrara, Via Saragat 1, 44100 Ferrara, Italy}
\affiliation{Istituto Nazionale di Fisica Nucleare, Laboratori Nazionali di Legnaro, \\ Via dell'Universit\`a 2 - 35020 Legnaro, Padova, Italy}
\affiliation{Istituto Nazionale di Fisica Nucleare, Sezione di Ferrara, Via Saragat 1, 44100 Ferrara, Italy}
\author{			G.L.\ Fogli}
\affiliation{Dipartimento Interateneo di Fisica ``Michelangelo Merlin,'' Via Amendola 173, 70126 Bari, Italy}
\affiliation{Istituto Nazionale di Fisica Nucleare, Sezione di Bari,  Via Orabona 4, 70126, Bari, Italy}
\author{	          E.\ Lisi}
\affiliation{Istituto Nazionale di Fisica Nucleare, Sezione di Bari,  Via Orabona 4, 70126, Bari, Italy}
\author{	          F.\ Mantovani}
\affiliation{Dipartimento di Fisica, Universit\`a di Ferrara, Via Saragat 1, 44100 Ferrara, Italy}
\affiliation{Istituto Nazionale di Fisica Nucleare, Sezione di Ferrara, Via Saragat 1, 44100 Ferrara, Italy}
\author{			A.M.\ Rotunno}
\affiliation{Dipartimento Interateneo di Fisica ``Michelangelo Merlin,'' Via Amendola 173, 70126 Bari, Italy}

%\date{{\today}}

\begin{abstract}
The KamLAND and Borexino experiments have observed, each at $\sim 4\sigma$ level, signals of
electron antineutrinos produced in the decay chains of thorium  and uranium in the Earth's crust and mantle
(Th and U geoneutrinos). Various pieces of geochemical and geophysical information allow an estimation of the crustal
geoneutrino flux components with relatively small uncertainties. The mantle component may then be inferred  
by subtracting the estimated crustal flux from the measured total flux.
To this purpose, we analyze in detail the experimental Th and U geoneutrino 
event rates in KamLAND and Borexino, 
including neutrino oscillation effects. We estimate the crustal flux at the two detector sites, 
using state-of-the-art information about the Th and U distribution on global and local scales. 
We find that crust-subtracted signals show hints of a residual mantle component, emerging
at $\sim\!2.4\sigma$ level by combining the KamLAND and Borexino data.
The inferred mantle flux slightly favors scenarios with relatively high Th and U abundances,
within $\pm 1\sigma$ uncertainties comparable to the spread of predictions from recent mantle models. 
\end{abstract}
\maketitle

%%%%%%%%%%%%%%%%%%%%%%%%%%%%%%%%%%%%%%%%%%%%%%%%%%%%%%%%%%%%%%%%%%%%%%
%%%% Section I %%%%%%%%%%%%%%%%%%%%%%%%%%%%%%%%%%%%%%%%%%%%%%%%%%%%%%%
%%%%%%%%%%%%%%%%%%%%%%%%%%%%%%%%%%%%%%%%%%%%%%%%%%%%%%%%%%%%%%%%%%%%%%

\section{Introduction}
\vspace*{-2mm}

The decay chains of uranium (U), thorium (Th), and potassium (K) in the Earth's interior
provide intense sources of terrestrial heat and, at the same time,
of low-energy electron antineutrinos ($\overline \nu_e$) --- the so-called geoneutrinos \cite{Fi07}.
Geoneutrinos from Th and U (but not from K) decay are detectable via the inverse beta decay (IBD) reaction,
\begin{equation}
\label{IBD} \overline\nu_e+ p\to n + e^+ \ (E_\nu>1.806\ \mathrm{MeV})\ ,
\end{equation}
and have recently been observed at $\sim 4\sigma$ level both in the  
KamLAND (KL) \cite{KL11} and in the Borexino (BX) \cite{BX10} experiments. 
The KL and BX measurements represent first steps
of a long-term research program which, by bridging particle physics and Earth science, will
provide unique clues on fundamental geophysical and geochemical issues \cite{Mc08,Mare,Hawa}.

Indeed, the geoneutrino flux and its spectrum encode relevant information about the
distribution of radiogenic elements
in the crust and in the mantle, which are thought to be the main Th and U reservoirs \cite{Mc08}.%
%-----------------------
\footnote{ Geochemical arguments disfavor significant amounts of Th and U in the Earth's core, see \protect\cite{Mc08}
and refs.\ therein. If the core were (hypothetically) a geoneutrino source, the ``mantle'' fluxes estimated in this work should be
interepreted as ``core+mantle'' fluxes.}
%-----------------
  In particular,  the total $\overline\nu_e$ flux probes
the total amount of radiogenic elements in the Earth,  while the energy spectrum is  
sensitive to the different Th and U components \cite{Fi07}. In principle, 
the angular spectrum (not yet experimentally accessible \cite{Baty}) can probe
the different mantle and crust source geometry \cite{Hoch}. Repeating such measurements 
at different locations can thus help to distinguish the site-dependent crustal components from 
the (approximately) site-independent mantle component of the flux, which
can also be more directly probed at oceanic sites \cite{Dy10}. 

Extracting such information is not straightforward,
since the geoneutrino flux represents a volume integral over Th and U abundances,
weighted by the inverse square distance, and modulated by the IBD cross section
and $\overline \nu_e$ oscillation probability (see \cite{Fi07} for details).
While the latter two ingredients are known with good accuracy, the
volume distribution of Th and U is subject to relatively large uncertainties, especially in the mantle \cite{Mc08}.
Therefore, in order to disentangle interesting pieces 
of information from particle physics data (geoneutrinos), one needs an interdisciplinary approach, 
including supplementary constraints or assumptions from Earth science (geophysics and geochemistry). 
For instance, in order to constrain the radiogenic heat flux, one may exploit its 
(partly model-dependent) covariances with the total geoneutrino flux \cite{Fi07,Cova}; or, in order to 
probe meteoritic expectations for the Th/U abundance ratio \cite{Rati}, one may assume that KL and BX 
experiments probe the same average Th/U in first approximation \cite{Prev}.

In this work we apply such an interdisciplinary approach to infer the mantle component of the geoneutrino flux,
which we obtain by subtracting accurately estimated crust components from the total measured fluxes.
In particular, concerning particle physics data,
we perform a detailed analysis of the total Th and U geoneutrino fluxes 
measured in KL and BX, including oscillation effects (Sec.~II). Concerning Earth science data,
we estimate the different crustal flux components in the two experiments, 
using state-of-the-art geochemical and geophysical information about the crust,
on both global and local scales (Sec.~III). The mantle component in KL and Borexino 
is then obtained by subtraction ($ \mathrm{mantle} = \mathrm{total} - \mathrm{crust}$). 
Within the reasonable assumption of site-independent mantle flux, 
the KL and BX results can be combined, yielding a mantle signal at $>2\sigma$ (Sec.~IV).
In comparison with current models of the mantle, the signal best fit
favors scenarios with relatively high Th and U mantle abundances, within
$\pm1\sigma$ uncertainties comparable with the spread of model predictions (Sec.~V).
The main results are summarized in Sec.~VI.
Statistical details and side results of our analysis are confined 
to Appendix A and B, respectively.

\vspace*{-3mm}
\section{Particle physics input: Analysis of KL and BX data} 
\vspace*{-2mm}

In this Section we update the KL and BX geoneutrino data analysis discussed in \cite{Prev},
by including recent constraints from world neutrino oscillation data as in \cite{Glob}, and
the latest KL geoneutrino data from \cite{KL11}. With respect to  \cite{Prev}, 
the BX data \cite{BX10} are the same, while the updated KL data are significantly more accurate. 
We use the KL energy spectra and detection efficiencies for $\overline\nu_e$ events as shown 
in Fig.~1 of \cite{KL11}, within the same statistical approach discussed in \cite{Prev}. 
In particular, the fit to KL and BX data involves a 7-dimensional manifold, 
%...................
\begin{equation}
\label{PAR} 
\mathrm{Parameters}=\{\delta m^2,\,\theta_{12},\,\theta_{13}
;\,R(\mathrm{Th})_\mathrm{KL}
,\,R(\mathrm{U})_\mathrm{KL}
,\,R(\mathrm{Th})_\mathrm{BX}
,\,R(\mathrm{U})_\mathrm{BX}
\}\ .
\end{equation}
%.................
where the
four $R$'s represent the KL and BX event rates from Th and U geoneutrinos,
expressed in Terrestrial Neutrino Units (1 TNU = $10^{-32}$~events per target proton per year).
It is useful to remind that  a reference (oscillated) $\overline\nu_e$ flux  
$\phi=10^6/$cm$^2/$s generates $R(\mathrm{Th})= 4.04$~TNU and 
$R(\mathrm{U})=  12.8$~TNU, 
while a  mass abundance ratio $\mathrm{Th}/\mathrm{U}=a(\mathrm{Th})/a(\mathrm{U})$ corresponds to   
$R(\mathrm{Th})/R({\mathrm{U}})=0.0696\cdot \mathrm{Th}/\mathrm{U}$ for a homogeneous source \cite{Fi07}.

In Eq.~(\ref{PAR}), the mass-mixing oscillation parameters $(\delta m^2,\,\theta_{12},\, \theta_{13})$ 
govern the flavor survival probability $P_{ee}$ of
both geo-$\overline\nu_e$ and background reactor $\overline\nu_e$, 
%..............................................
\begin{equation}
\label{Pee}
P_{ee} = P(\overline\nu_e\to\overline\nu_e) = \cos^4\theta_{13}\left(1-\sin^22\theta_{12}\sin^2\left(\frac{\delta m^2 L}{4\,E}\right)\right)+\sin^4\theta_{13} \ ,
\end{equation}
%...............................................
$L$ and $E$ being the $\overline\nu_e$ path length  and energy, respectively, in natural units. 
As in \cite{Prev}, we make the reasonable approximation of oscillation-averaged $P_{ee}$ for geoneutrinos, 
\begin{equation}
\label{Pave}
\langle P_{ee}\rangle \simeq \cos^4\theta_{13}\left(1-\frac{1}{2}\sin^22\theta_{12}\right)+\sin^4\theta_{13}\ . 
\end{equation}
We have verified a posteriori that, within current uncertainties,  removing this simplifying assumptions does not spoil our main results; see 
Appendix~A for details.  

For $\langle P_{ee}\rangle$, we adopt the reference $1\sigma$ ranges $\sin^2\theta_{12}\simeq 0.306\pm 0.017$ and 
$\sin^2\theta_{13}\simeq 0.021\pm 0.007$ from the global analysis of oscillation data (from solar, atmospheric, accelerator, and reactor
$\nu$ experiments) in \cite{Glob}, implying:
%...................................
\begin{equation}
\label{Pbest}
\langle P_{ee}\rangle \simeq 0.551 \pm 0.015 \ (1\sigma)\ .
\end{equation}
%......................................
The $3\%$ fractional uncertainty of $\langle P_{ee}\rangle$  is much smaller than other (mainly experimental) errors
affecting current geoneutrino event rates. In any case, it is 
properly taken into account by marginalization in the global $\chi^2$ fit of oscillation plus geoneutrino data, 
which yields measured event rates $R$ which consistently include {\em all\/} the uncertainties from particle physics inputs.%
%----------------------------
\footnote{After our work was completed, high-precision
measurements of $\sin^2\theta_{13}$ were announced by two reactor neutrino experiments: Daya Bay  
($\sin^2\theta_{13}=0.024 \pm 0.004$) \protect\cite{Daya} and RENO  ($\sin^2\theta_{13}=0.029 \pm 0.006$) \protect\cite{RENO}.   
The currently recommended $1\sigma$ range from world reactor data, $\sin^2\theta_{13}\simeq 0.025 \pm 0.003$ \protect\cite{PDGB}, is fully contained
in the $1\sigma$ range used herein: $\sin^2\theta_{13}\simeq 0.021\pm 0.007$ \protect\cite{Glob}. An updated global
analysis of the neutrino mass-mixing parameters, including Daya Bay, RENO and other recent oscillation data, is reported elsewhere \protect\cite{Nu12}. 
In any case, such update would have negligible effects on the geoneutrino results presented in this work.} 
%-------------------------

Figure~1 shows the results of our analysis of the total (Th+U) geoneutrino event rate $R$ in KL and BX, in terms
of standard deviations ($N_\sigma=\sqrt{\Delta\chi^2 }$) from the best fit. 
The null hypothesis of vanishing geoneutrino flux  ($R=0$) is rejected at $N_\sigma\simeq 4.2$ by both KL and BX, 
in good agreement with the corresponding official results in \cite{KL11} and \cite{BX10}; see also Appendix B.
Note that the KL and BX curves are not linear and symmetric, as it would be the case for gaussian errors: 
indeed, the asymmetry and nonlinearity increase for decreasing event rate $R$ (especially for the low-statistics BX data), as a result
of Poisson fluctuations, whose statistics is properly accounted for in our $\chi^2$ analysis \cite{Prev}. As a consequence,
the current KL and BX event rates cannot be simply summarized in terms of central values and $\pm 1\sigma$ errors: 
non-gaussian uncertainties must be properly taken into account, in order 
to exploit the full potential of the available geoneutrino data.

%%%%%%%%%%%%%%%%%%%%%%%%%%% FIGURE 1 %%%%%%%%%%%%%%%%%%%%%%%%%%%%%%%%
\begin{figure}[t]
\centering
\includegraphics[width=0.40\columnwidth]{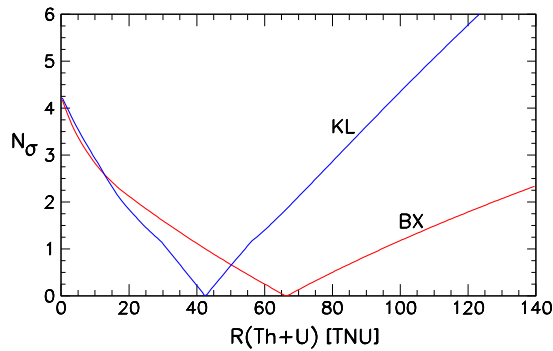}
\caption{Constraints on the total geoneutrino event rate in KL and BX, in terms of standard deviations $N_\sigma$ from the best fit.
\label{fig1}}
\end{figure}
%%%%%%%%%%%%%%%%%%%%%%%%%%%%%%%%%%%%%%%%%%%%%%%%%%%%%%%%%%%%%%%%%%%%%%

In order not to loose precious information, in the following we shall mostly refer to the 
full data analysis in terms of separate (not summed) Th and U event rates. 
Figure~2 shows, in particular, the main results of our analysis of KL and BX data, 
in the plane charted by the geoneutrino event rates $R(\mathrm{U})$ and $R(\mathrm{Th}$) for KL (upper panel) and BX (lower panel). 
In both panels, the thick dot represents the best fit,
while the curves correspond to the $n\sigma$ contours ($\Delta\chi^2=n^2$, for $n=1$, 2 and 3), whose projections 
provide the $n\sigma$ bounds for the corresponding parameter \cite{PDGB}. Note that, in Fig.~2, the $1\sigma$ contour for KL 
is closed, i.e., KL can separate the Th and U components at the (weak) level of $\sim 1\sigma$, due to higher
statistics and better spectral information.
However, already at $2\sigma$, the KL and the BX contours are no longer closed; indeed,  
the strong anticorrelation of the $n\sigma$ isolines reflects the fact that the
KL and BX spectra are currently more sensitive to the total Th+U flux than to their separate Th and U components which, to some extent, 
can be traded one for the other. Figure~2 also anticipates schematically the subtraction of 
the estimated crustal rates
(with their associated $\pm 3\sigma$ errors), as discussed  in detail in the next Sections.

%%%%%%%%%%%%%%%%%%%%%%%%%%% FIGURE 2 %%%%%%%%%%%%%%%%%%%%%%%%%%%%%%%%
\begin{figure}[b]
\centering
\includegraphics[width=0.48\columnwidth]{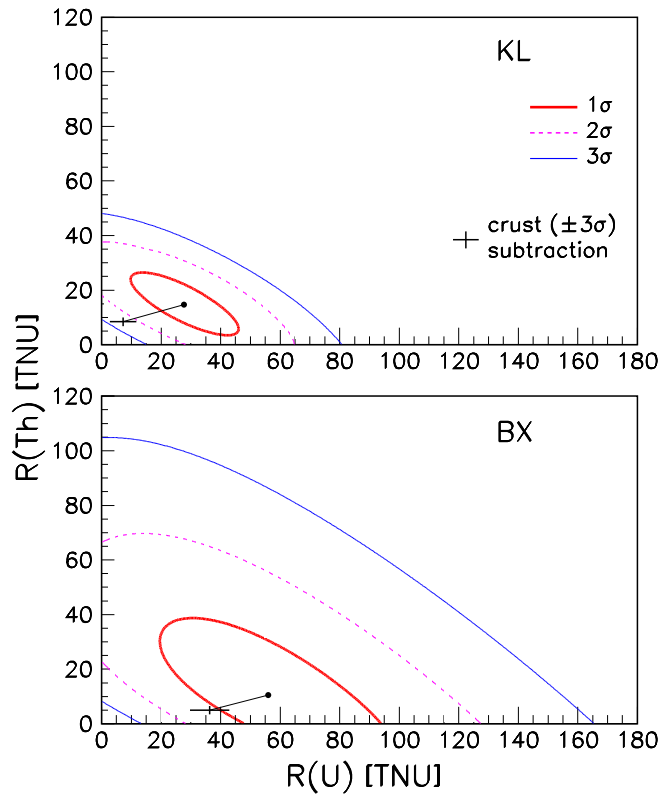}
\caption{Results of our analysis of geoneutrino rates $R(\mathrm{Th})$ and $R(\mathrm{U})$
expressed in TNU, for both KL (upper panel) and BX (lower panel). The curves represent the $n\sigma$
contours ($\Delta\chi^2=n^2$) around the best fit (thick dot). Also shown is the 
shift of the best fit points after subtraction of the estimated crustal components (with their $\pm 3\sigma$ errors).
See the text for details. 
\label{fig2}}
\end{figure}
%%%%%%%%%%%%%%%%%%%%%%%%%%%%%%%%%%%%%%%%%%%%%%%%%%%%%%%%%%%%%%%%%%%%%%

\section{Earth science input: Crustal flux estimates in KL and BX}

In order to estimate the crustal geoneutrino flux we need 
a global model for the Earth crust and a sufficiently detailed model 
for the local contribution. Indeed, due to the inverse square law for the flux, 
the crust portions within and outside a radius of $O(500)$~km from the detector
provide comparable flux contributions in both KL and BX \cite{Fi07}. 
In particular, for both KL and BX, we use an accurate description of the local crust
extending over $\sim\!2.5\times 10^5$~km$^2$ and down to 30--40 km of depth (Moho
surface), which contributes to $\sim\!40\%$ of the total geo-$\nu$ signal. 
For farther portions of the crust, a coarser description in terms 
of $2^\circ\times2^\circ$ tiles is sufficient.  

In this section we discuss global and local properties of the crust, building
upon previous works on the subject \cite{Fi07,KLOC,Colt}. 
In particular,
we report our calculation of the crustal Th and U fluxes at KL and BX, 
with estimated $1\sigma$ uncertainties of O(10\%). It should be noted, however, that precise estimates
for such uncertainties play no significant role in this work: even if the crustal flux errors 
were all conservatively doubled or tripled, the final results for the mantle signal
would only change by a tiny fraction of one standard deviation (see Appendix~A). 

\subsection{Global model of the crust} 
 
Our global model for the crust is based on a geophysical $2^\circ\times2^\circ$ tiled map \cite{Tile}, as well as
on the Th and U mass abundances recommended in \cite{Sedi} for sedimentary layers and in \cite{Rudn} for 
the upper, middle, and lower crust.  For the lower crust, the values in the literature encompass a large interval: we adopt a mean value together with an uncertainty indicative of the spread of published values \cite{Fi07}.
After removal of a tiny portion of crust extending for a few hundred km around each site 
(local or ``LOC'' contribution, as defined in the next subsections), the global model is used to evaluate the 
remaining geoneutrino flux at both KL and BX (rest-of-the-crust or ``ROC'' contribution). 

Table~I reports the input total mass and adopted values and $\pm 1\sigma$ errors 
for radiogenic abundances in each global reservoir, 
as well as the output ROC event rates in KL and BX, 
multiplied by the central value of
the probability in Eq.~(\ref{Pbest}). [The $P_{ee}$ errors are already accounted for in
the experimental data fits of Figs.~1~and~2.]
Total  errors in the last row are obtained by summing partial errors in quadrature.

%%%%%%%%%%%%%%%%%%%%%%%%%%%%%%%%%%%%%%%%%%%%%%%%%%%%%%%%%%%%%%%%%%%%%%%%%%%%%%%%%%%%%%%%%%%%%%%%%%%%%%%%%%%%
\begin{table}[h]
\begin{minipage}{\textwidth}
\caption{\label{GLOBAL} Inputs and outputs of the global model of the crust adopted in this work. The first four columns
report, for each reservoir, its mass $M$ and the adopted Th and U mass abundances. The last four columns 
report  the estimated rest-of-the-crust (ROC)
event rates, as obtained by excluding from the total crust the local (LOC) portions defined in the text. Quoted 
errors are at $1\sigma$.}
\begin{ruledtabular}
\begin{tabular}{cccccccc}
%=============================================================================================================================================================
          		&                   	&               	&      	&\multicolumn{2}{c}{ROC rates for KL}   	&\multicolumn{2}{c}{ROC rates for BX}   \\
Reservoir 		& $M$ [$10^{22}$~kg]	& $a$(Th) [$\mu$g/g] 	& $a$(U) [$\mu$g/g] 	& $R$(Th) [TNU] 	& $R$(U) [TNU] 	& $R$(Th) [TNU]	& $R$(U) [TNU] \\
\hline%-------------------------------------------------------------------------------------------------------------------------------------------------------
Sediments		& 	0.11				& $6.9\pm0.8$ 	& $1.7\pm0.2$ 	& $0.10\pm0.01$	& $0.34\pm0.04$ 	& $0.22\pm0.03$ 	& $0.82\pm0.09$ \\
Upper crust		& 	0.70				& $10.5\pm1.0$ 	& $2.7\pm0.6$ 	& $0.99\pm0.10$	& $3.64\pm0.80$ 	& $1.66\pm0.16$ 	& $6.42\pm1.43$  	\\
Middle crust		& 	0.71				& $6.5\pm0.5$ 	& $1.3\pm0.4$ 	& $0.62\pm0.05$	& $1.80\pm0.56$ 	& $1.11\pm0.09$ 	& $3.32\pm1.02$  	\\
Lower crust
		       	& 	0.66				& $3.7\pm2.4$ 	& $0.6\pm0.4$ 	& $0.34\pm0.22$	& $0.80\pm0.54$ 	& $0.59\pm0.39$ 	& $1.44\pm0.96$  	\\
Oceanic crust\footnotemark[1]
 				& 	0.60				& $0.22\pm0.07$ 	& $0.10\pm0.03$ 	& $0.02\pm0.01$	& $0.11\pm0.04$ 	& $0.01\pm.003$	& $0.07\pm0.02$  	\\
\hline%-------------------------------------------------------------------------------------------------------------------------------------------------------
Total			&	   				& 				& 			 	& $2.07\pm0.25$	& $6.71\pm1.12$ 	& $3.72\pm0.43$ 	& $12.07\pm2.00$  \\
\end{tabular}
\end{ruledtabular}
%\footnotetext[1]{For the lower crust, uncertainties are indicative of the spread of published values.}
\footnotetext[1]{For the oceanic crust, uncertainties are taken from  private communication with R.~Rudnick.}
%=============================================================================================================================================================
\end{minipage}
\end{table}
%%%%%%%%%%%%%%%%%%%%%%%%%%%%%%%%%%%%%%%%%%%%%%%%%%%%%%%%%%%%%%%%%%%%%%%%%%%%%%%%%%%%%%%%%%%%%%%%%%%%%%%%%%%%

From the global model we estimate the following contributions to the
radiogenic crustal heat: $H(\mathrm{Th})=4.02\pm 0.47$~TW
and $H(\mathrm{U})=3.40\pm0.56$~TW. Assuming a mass ratio $\mathrm{K}/\mathrm{U} \sim 13000$ \cite{Arev},
the additional K contribution would be $H(\mathrm{K})\simeq 1.5\pm 0.2$~TW and
the total estimated crustal heat would amount to
$H(\mathrm{Th}+\mathrm{U}+\mathrm{K})\simeq 8.9\pm 1.2$~TW, where errors
have been added linearly, due to the high positive correlations among the three
radiogenic element abundances.

\vspace*{-4mm}
\subsection{Local model of the crust around KL}
\vspace*{-2mm}

The local crust at the KL site (Kamioka)
is defined in terms of six $2^\circ\times2^\circ$ tiles, 
supplemented with geochemical information on a $0.25^\circ\times0.25^\circ$ grid and on a detailed
map of the crust depth \cite{KLOC}. The possible effects of the subducting slab beneath Japan are
considered, and the uncertainties arising from the debated (continental or oceanic) nature of
the crust below the Japan Sea are also taken into account. The maximal and minimal excursions
of various inputs and uncertainties are taken as a proxy for the $\pm 3\sigma$ error range.

Table~II summarizes the estimated LOC contributions to the geoneutrino signal in KL, together with
their estimated $1\sigma$ errors; total errors are obtained by summing in quadrature. 
Further details can be found in \cite{Fi07,KLOC}.

%%%%%%%%%%%%%%%%%%%%%%%%%%%%%%%%%%%%%%%%%%%%%%%%%%%%%%%%%%%%%%%%%%%%%%%%%%%%%%%%%%%%%%%%%%%%%%%%%%%%%%%%%%%%%%%%%%%%%%%%%%%%%%%%%%%%%%%%%%%%%%%%%%%%%%%%%%%%%%
\begin{table}[t]
\begin{minipage}{\textwidth}
\caption{\label{LOCKL} Local (LOC) contributions to the geoneutrino signal in KL. Quoted errors are at $1\sigma$.}
\begin{ruledtabular}
\begin{tabular}{ccc}
%=============================================================================================================================================================
Reservoir 		& $R$(Th) [TNU] 	& $R$(U) [TNU] \\
\hline%-------------------------------------------------------------------------------------------------------------------------------------------------------
Six tiles\footnotemark[1]
				& $3.20\pm0.37$ & $11.17\pm 0.65$	\\
Subducting slab	& $0.90\pm0.27$ & $2.02\pm 0.61$ \\
Japan sea		& $0.09\pm0.03$ & $0.34\pm 0.10$ \\
\hline%-------------------------------------------------------------------------------------------------------------------------------------------------------
LOC total		& $4.19\pm0.46$ & $13.53\pm 0.90$
\end{tabular}
\end{ruledtabular}
\footnotetext[1]{The six-tiles errors include uncertainties on the crust composition, depth, and map discretization.}
%=============================================================================================================================================================
\end{minipage}
\end{table}
%%%%%%%%%%%%%%%%%%%%%%%%%%%%%%%%%%%%%%%%%%%%%%%%%%%%%%%%%%%%%%%%%%%%%%%%%%%%%%%%%%%%%%%%%%%%%%%%%%%%%%%%%%%%%%%%%%%%%%%%%%%%%%%%%%%%%%%%%%%%%%%%%%%%%%%%%%%%%%

\vspace*{-4mm}
\subsection{Local model of the crust around BX} 
\vspace*{-2mm}

The local crust at the BX site (Gran Sasso)
is defined in terms of a $2^\circ\times2^\circ$ central tile (CT) and of the
rest of the region (RR) formed by the surrounding six tiles minus the CT \cite{Colt}. Geophysical
features of the local crust (geometry, density, seismic velocities etc.) are
reported in \cite{Colt}. From a geochemical viewpoint, the CT sedimentary is a mixture
of four main reservoirs, which have been probed by direct sampling of Th and U
abundances. In the upper and lower crust one can recognize two components
(felsic and mafic rocks) which are also probed by direct measurements. 
Average elemental abundances for the two groups were calculated and seismic arguments used 
in order to fix their relative amounts within the upper crust. In the lower
crust, the fraction of felsic and mafic rocks was estimated on the basis
of geophysical and geochemical information. The same Th and U abundances were assumed 
in the CT and in the RR. The maximal and minimal excursions
of various input values and uncertainties were taken as a proxy for the $\pm 3\sigma$ error range.

Table~III summarizes the estimated LOC contributions to the geo-$\nu$ signal in BX, together with
their $1\sigma$ errors; total errors are obtained by summing in quadrature. Adopted
abundances are also reported.  See \cite{Colt} for details.

%%%%%%%%%%%%%%%%%%%%%%%%%%%%%%%%%%%%%%%%%%%%%%%%%%%%%%%%%%%%%%%%%%%%%%%%%%%%%%%%%%%%%%%%%%%%%%%%%%%%%%%%%%%%
\begin{table}[h]
%\begin{minipage}{\textwidth}
\caption{\label{LOCBX} Local (LOC) abundances and contributions to the geoneutrino signal in BX. Quoted errors are at $1\sigma$.}
\begin{ruledtabular}
\begin{tabular}{ccccc}
%=============================================================================================================================================================
Reservoir 		& $a(\mathrm{Th})$ [$\mu$g/g] & $a(\mathrm{U})$ [$\mu$g/g] & $R$(Th) [TNU] & $R$(U) [TNU] \\
\hline%-------------------------------------------------------------------------------------------------------------------------------------------------------
Sediments		& $2.00\pm0.17$ & $0.80\pm0.07$ & $0.40\pm0.04$ & $2.53\pm 0.21$	\\
Upper crust		& $8.1\pm1.6$   & $2.20\pm0.43$ & $1.21\pm0.24$ & $4.94\pm 0.96$ \\
Lower crust		& $2.6\pm 1.2$  & $0.30\pm0.10$ & $0.25\pm0.11$ & $0.34\pm 0.11$ \\
\hline%-------------------------------------------------------------------------------------------------------------------------------------------------------
LOC total		&               &               & $1.86\pm0.27$ & $7.81\pm 0.99$
\end{tabular}
\end{ruledtabular}
%=============================================================================================================================================================
%\end{minipage}
\end{table}
%%%%%%%%%%%%%%%%%%%%%%%%%%%%%%%%%%%%%%%%%%%%%%%%%%%%%%%%%%%%%%%%%%%%%%%%%%%%%%%%%%%%%%%%%%%%%%%%%%%%%%%%%%%%

\vspace*{-5mm}
\subsection{Estimated crustal geoneutrino rates and uncertainties at KL and BX}
\vspace*{-2mm}

Table~IV summarizes the results of this section, in terms of LOC, ROC and total
event rates in KL and BX, with errors added in quadrature. The fractional 
uncertainties, all of $O(10\%)$, are not crucial in the context of our analysis, which is dominated
by experimental errors (see below and Appendix~A). In the future, however, it might be useful to address some geophysical 
uncertainties---such as those related to the 
global crust thickness---which have been neglected herein, in comparison with the larger geochemical abundance uncertainties.

Concerning the central values, it should be noted that
the total estimated rates in BX are lower than in KL, 
contrary to previous estimates and expectations (based on the fact that BX is surrounded by 
thicker crust than KL) \cite{Fi07,Hawa}. This counterintuitive result is mainly driven by the improved description 
of the local BX crust performed in \cite{Colt}, leading to a significantly deeper sediment layer,  
and to a depletion of Th and U in all local crust reservoirs, as compared to 
previous estimates using no (or less accurate) local models.

%%%%%%%%%%%%%%%%%%%%%%%%%%%%%%%%%%%%%%%%%%%%%%%%%%%%%%%%%%%%%%%%%%%%%%%%%%%%%%%%%%%%%%%%%%%%%%%%%%%%%%%%%%%%
\begin{table}[h]
%\begin{minipage}{\textwidth}
\caption{\label{TOTAL} Summary of LOC, ROC and total crust contributions to the geo-$\nu$ signal in KL and BX. Quoted errors are at $1\sigma$.}
\begin{ruledtabular}
\begin{tabular}{ccccc}
%=============================================================================================================================================================
          		&\multicolumn{2}{c}{KL event rates}   	&\multicolumn{2}{c}{BX event rates}   \\
Reservoir 		& $R$(Th) [TNU] 	& $R$(U) [TNU] 			& $R$(Th) [TNU]	& $R$(U) [TNU] \\
\hline%-------------------------------------------------------------------------------------------------------------------------------------------------------
LOC				& $4.19\pm0.46$ & $13.53\pm0.90$ & $1.86\pm0.27$ & $7.81\pm 0.99$	\\
ROC				& $2.07\pm0.25$ & $6.71\pm1.12$  & $3.60\pm0.43$ & $12.07\pm 2.00$ \\
\hline%-------------------------------------------------------------------------------------------------------------------------------------------------------
Crust total		& $6.26\pm0.52$ & $20.24\pm1.43$ & $5.46\pm0.51$ & $19.88\pm2.24$
\end{tabular}
\end{ruledtabular}
%=============================================================================================================================================================
%\end{minipage}
\end{table}
%%%%%%%%%%%%%%%%%%%%%%%%%%%%%%%%%%%%%%%%%%%%%%%%%%%%%%%%%%%%%%%%%%%%%%%%%%%%%%%%%%%%%%%%%%%%%%%%%%%%%%%%%%%%

\section{Crust subtraction and geoneutrino rates from the mantle} 

In the previous two sections we have determined, for both KL and BX, the experimental total rates $R$ (Fig.~1)
and the theoretical crustal rates $R_\mathrm{crust}$ (Table~IV) due to Th and U geoneutrinos, together with the
associated uncertainties. In this section we infer the mantle rates, by subtracting the estimated crustal components from the experimental
total rates,
%.........................................................................................
\begin{equation}
R(\mathrm{mantle}) = R(\mathrm{total,\ exp}) - R(\mathrm{crust,\ theo})\ .
\end{equation}
%.........................................................................................

Going back to Fig.~2, in each panel, crustal rate subtraction is graphically
shown as a shift of the best-fit point by a vector (ending with crossed
error bars) defined by the total Th and U crust rates and their $\pm3\sigma$ errors (see Table~IV).
In principle, such estimated crust uncertainties should be properly combined with 
the experimental rate uncertainties. However, the latter are currently larger by (more than) an order
of magnitude, and dominate the error associated to the subtraction procedure. 
Indeed, in Appendix~A we demonstrate that, in the present context,  crustal rate errors are practically
insignificant, even if they are conservatively
inflated by a factor of a few. As a consequence, the $n\sigma$ contours for the mantle
rates can be simply obtained by rigidly shifting the $n\sigma$ contours in Fig.~1, together with the
best fit point, along the slanted line. Numerically, this amounts to a simple translation of the two-dimensional
$\chi^2$ functions for both KL and BX.

%%%%%%%%%%%%%%%%%%%%%%%%%%% FIGURE 3 %%%%%%%%%%%%%%%%%%%%%%%%%%%%%%%%
\begin{figure}[t]
\centering
\includegraphics[width=0.3\columnwidth]{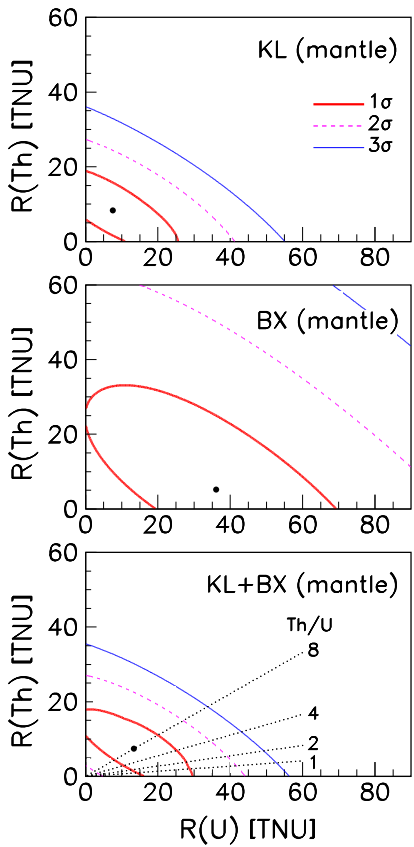}
\caption{Mantle geoneutrino rates obtained from crustal rate subtraction in 
KL (upper panel) and BX (middle panel), as well as in their combination KL+BX (lower panel). The curves represent the $n\sigma$
contours around the best fit. In the lower panel, lines of constant Th/U abundance 
ratio are also shown. The null hypothesis (no mantle signal) is disfavored at $>2\sigma$
by KL+BX. 
\label{fig3}}
\end{figure}
%%%%%%%%%%%%%%%%%%%%%%%%%%%%%%%%%%%%%%%%%%%%%%%%%%%%%%%%%%%%%%%%%%%%%%

Figure~3 shows the allowed $n\sigma$ regions of the mantle rates resulting from crust subtraction in KL (upper panel)
and in BX (middle panel). In both cases, positive values for the Th and U mantle rates are preferred. The fact that
the best fit rates are well within the physical region (and none of them is negative) 
is an encouraging sanity check of the crust subtraction procedure. 
More precisely, 
we find that the null hypothesis of no mantle signal (i.e., the origin of the axes) is disfavored  
at about $1.7\sigma$ level in KL and $2.0\sigma$ in BX. Moreover, the allowed
regions in KL and BX largely overlap. Therefore, we have obtained
two consistent hints at $\geq 1.7\sigma$ level
in favor of a geoneutrino signal coming from the mantle. 
Nonzero mantle fluxes were also suggested (but not quantified) by the analyses in \cite{KL11,Prev}.

The above hints can now be properly combined under the assumption of site-independent mantle flux, which
is justified in the absence of significant indications in favor of local mantle anomalies
below the KL or BX sites. (Possible geochemical mantle anomalies are still debated and, in any case,
they are strongly model dependent \cite{Anom}.)
 Under such hypothesis,     
the combination of KL and BX constraints on mantle geoneutrino rates 
amount to summing the corresponding $\chi^2$ functions. The results are shown in the lower
panel of Fig.~3. The mantle signal common to the KL and BX emerges now with greater statistical significance, the
null hypothesis being rejected at $2.4\sigma$ (about $98.4\%$ C.L.).
This result represents an encouraging first step towards 
a better understanding  of the mantle via geoneutrinos,
and can already provide valuable indications in 
comparison with various mantle models, as shown in the next Section.
 
Of course, the inferred mantle signal is not yet accurate enough to 
probe more detailed issues, such as the mantle Th/U ratio.
In particular, in Fig.~3, the KL+BX contour at $1\sigma$ appears to be  compatible
with any possible Th/U mantle ratio (some isolines being shown to guide the eye), so that
the preference for Th/U~$\simeq 8$ is not statistically significant.
A future reduction of the experimental errors 
(which is conceivable with longer exposures, better background rejection, and 
additional experiments) would be desirable to get a mantle
geoneutrino signal with greater impact for 
geophysics and geochemistry.

\vspace*{-2mm}
\section{Comparison with mantle models}
\vspace*{-2mm}

In this Section we compare the inferred mantle geoneutrino signal with predictions derived
from various published mantle models, hereafter referred to as: Turcotte and Schubert 2002 \cite{Turc},
Anderson 2007 \cite{Ande}, Palme \& O'Neill 2003 \cite{Palm}, Allegre et al.\ 1995 \cite{Alle}, McDonough \& Sun 1995 \cite{McDo},
Lyubetskaya \& Korenaga 2007 \cite{Lyub}, Javoy et al.\ 2010 \cite{Javo}. It should be remarked that,
in general, the 
models mainly focus on the ``primitive mantle'' (PM, with mass $M_\mathrm{PM}=4.03\times 10^{24}$~kg) prior to differentiation
into crust and ``present mantle,'' while our results correspond, of course, only to the present mantle.
Therefore, in order to perform a comparison with the results obtained in the previous
Section, for each model we remove from the PM the crustal Th and U masses (which, in our global
crust model of Sec.~III, amount to $M_\mathrm{Th}=15.30\pm 1.77$ 
and $M_\mathrm{U}=3.45\pm0.57$ at $1\sigma$, in units of $10^{16}$~kg). 

The distribution of the remainder Th and U masses within the present mantle volume is subject to a lively debate, 
opposing homogeneous versus inhomogeneous (e.g., layered) models \cite{Fi07,Mc08}.
For the purposes of this work, we consider two representative extreme scenarios, yielding
``high'' and ``low'' mantle geoneutrino rates. 
The ``high rate'' (homogeneous) scenario is obtained by
subtraction of the Th and U crustal masses at the lower end of their $1\sigma$ range, and 
distributing the remainder in the whole mantle at constant density. 
The ``low rate''(inhomogeneous)  scenario is obtained by subtracting from the PM
the Th and U crustal masses at the upper end of their $1\sigma$ range, and
placing all the remainder in the so-called D" layer (250~km thickness) just above the core-mantle 
boundary. In both cases,
averaged oscillations are included.
Note that the various models are based on different assumptions or input values about
the primitive chondritic material, which lead to further differences in the Th and U
contents and in the associated radiogenic heat in the present mantle.

Table~V summarizes our estimated ``low'' and ``high'' Th and U 
mantle geoneutrino rates as derived from each mantle model, together with
the associated total heat $H(\mathrm{Th}+\mathrm{U})$.  Note that two models
(Allegre et al.\ 1995,  McDonough \& Sun 1995) are practically identical for our purposes.
In one model (Javoy et al.\ 2010), characterized by rather low radiogenic abundances,
the Th content of the PM is slightly lower than the corresponding content of the crust in our ``low'' scenario;  
crustal subtraction would then yield slightly negative results, which have been just set to zero.
Since not all models in Table~V are endowed with PM error estimates, we do not
quote individual errors for the present mantle rates; their spread is, however, indicative 
of the large theoretical uncertainties typically associated to mantle observables.

%%%%%%%%%%%%%%%%%%%%%%%%%%%%%%%%%%%%%%%%%%%%%%%%%%%%%%%%%%%%%%%%%%%%%%%%%%%%%%%%%%%%%%%%%%%%%%%%%%%%%%%%%%%%
\begingroup
\squeezetable
\begin{table}[h]
%\begin{minipage}{\textwidth}
\caption{\label{MODELS} \scriptsize
Geoneutrino event rates derived from various models of the primitive mantle (PM), under different
assumptions about the Th and U distributions in the present mantle, leading to ``low'' and ``high''
rates. The first three columns characterize the 
original PM model in terms of Th and U masses. 
After crustal subtraction and redistribution of the remaining Th and U masses in the present mantle,
we derive the oscillated Th and U mantle event rates, the Th+U heat and the Th/U ratio as reported in the last eight columns,
for the ``low'' and ``high'' scenarios.}
\begin{ruledtabular}
\begin{tabular}{lcc|cccc|cccc}
%=============================================================================================================================================================
\multicolumn{3}{c|}{Primitive mantle  characteristics} & \multicolumn{4}{c|}{Present mantle, ``low'' scenario} &\multicolumn{4}{c}{Present mantle, ``high'' scenario}   \\
Model & $M_\mathrm{Th}$ & $M_\mathrm{U}$
                                 & $R(\mathrm{Th})$ & $R(\mathrm{U})$ & $H(\mathrm{Th+U})$ & Th/U &$R(\mathrm{Th})$ & $R(\mathrm{U})$ & $H(\mathrm{Th+U})$ & Th/U \\ 
      &[$10^{17}$kg]   & [$10^{17}$kg]  & [TNU]            & [TNU]           &      [TW]          &      & [TNU]            & [TNU]          &     [TW] & \\
\hline%-------------------------------------------------------------------------------------------------------------------------------------------------------
Turcotte \& Schubert 2002     & 3.62 & 0.90 & 2.7           & 9.8            & 17.0        & 3.9        & 3.9               & 14.7           & 19.0     & 3.8 \\ 
Anderson 2007                 & 3.13 & 0.78 & 2.3           & 8.4             & 14.5        & 3.9        & 3.4              & 12.8           & 16.6     & 3.8 \\ 
Palme \& O'Neil 2003          & 2.06 & 0.54 & 1.3           & 5.7             & 9.1         & 3.4        & 2.1              &  9.2           & 11.2     & 3.4 \\ 
Allegre et al.\ 1995          & 1.80 & 0.46 & 1.1           & 4.7             & 7.7         & 3.6        & 1.9              &  8.0           & 9.8      & 3.5 \\ 
McDonough \& Sun 1995         & 1.80 & 0.46 & 1.1           & 4.7             & 7.7         & 3.6        & 1.9              &  8.0           & 9.8      & 3.5 \\
Lyubetskaya \& Korenaga 2007  & 1.26 & 0.34 & 0.7           & 3.3             & 5.0         & 2.0        & 1.2              &  6.0           & 7.0      & 3.0 \\
Javoy et al.\ 2010            & 0.48 & 0.14 & 0.0           & 1.0             & 0.8         & 0.0        & 0.4              &  3.0           & 2.8      & 1.7 \\ 
\end{tabular}
\end{ruledtabular}
%=============================================================================================================================================================
%\end{minipage}
\end{table}
\endgroup
%%%%%%%%%%%%%%%%%%%%%%%%%%%%%%%%%%%%%%%%%%%%%%%%%%%%%%%%%%%%%%%%%%%%%%%%%%%%%%%%%%%%%%%%%%%%%%%%%%%%%%%%%%%%

%%%%%%%%%%%%%%%%%%%%%%%%%%% FIGURE 4 %%%%%%%%%%%%%%%%%%%%%%%%%%%%%%%%
\begin{figure}[t]
\centering
\includegraphics[width=0.62\columnwidth]{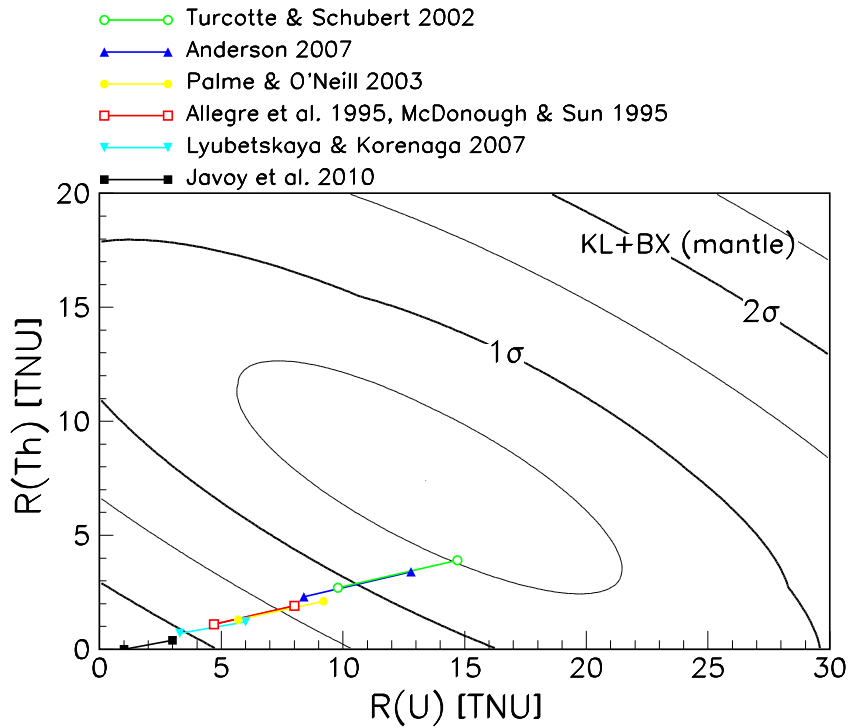}
\caption{Comparison of experimental constraints and model predictions in the plane charted by the Th and U mantle rates.
Each model leads to extremal case of ``low'' and ``high'' rates, connected by lines to guide the eye. The KL+BX constraints
are shown as $n\sigma$ contours in steps of $0.5\sigma$. See the text for details.
\label{fig4}}
\end{figure}
%%%%%%%%%%%%%%%%%%%%%%%%%%%%%%%%%%%%%%%%%%%%%%%%%%%%%%%%%%%%%%%%%%%%%%

Figure~4 shows the same KL+BX mantle rate constraints as in the lower panel  
of Fig.~3, but with superimposed model  predictions from Table~V. In order to guide
the eye, the ``low'' and ``high'' predictions for each model
are connected by a straight line; moreover, the KL+BX $n\sigma$ isolines are shown
in steps of $0.5\sigma$. 
It appears that the data prefer mantle models
with relatively high radiogenic contents (e.g., Turcotte \& Schubert 2002) 
and disfavor at $\sim 2\sigma$ those with low contents (e.g., Javoy et al.\ 2010).
This indication is rather interesting and deserves to be checked and investigated with
further data, given its potential impact on mantle model building.   

The near alignment of all model predictions in Fig.~4 reflects the narrowness of
the Th/U mantle ratio in Table~V, $\mathrm{Th/U}\in[1.7,\,3.9]$. In this context, it makes sense 
to marginalize away the Th/U ratio within such range, and to express the resulting rates 
in terms of the total (Th+U) event rate. We find at $1\sigma$ that 
%.........................
\begin{equation}
\label{Range}
R(\mathrm{Th+U,\ mantle})\simeq 23\pm 10~\mathrm{TNU}\  (\mathrm{KL+BX}, \mathrm{for\ Th/U}\in[1.7,\,3.9])\ ,
\end{equation}
%............................
with nearly gaussian error distribution (not shown). At present, this represents our best estimate for the mantle geoneutrino flux,
as derived by using inputs from particle physics (KL, BX, and oscillation data)
and from Earth sciences (crustal data and mantle Th/U ratio).

Finally, Fig.~5 shows a comparison between theory and data in terms of the
Th+U mantle rate (in TNU) and radiogenic heat (in TW). The various model
predictions, shown as lines connecting the ``low'' and ``high'' cases in Table~V, 
can be compared to the mantle rate inferred in Eq.~(\ref{Range}) and shown as a horizontal
$1\sigma$ band. Notice that the $1\sigma$ experimental error is already comparable
to the spread of the theoretical rate predictions: therefore, improvements
in the experimental accuracy by a factor of two or more, would allow a statistically
significant model discrimination.
At present we note that, at $1\sigma$, the data favor
models with relatively high mantle heat $H(\mathrm{Th+U})$, such as Anderson 2007
and Turcotte \& Schubert 2002. However, no model can be really excluded at
$\sim 2\sigma$.

%%%%%%%%%%%%%%%%%%%%%%%%%%% FIGURE 5 %%%%%%%%%%%%%%%%%%%%%%%%%%%%%%%%
\begin{figure}[h]
\centering
\includegraphics[width=0.6\columnwidth]{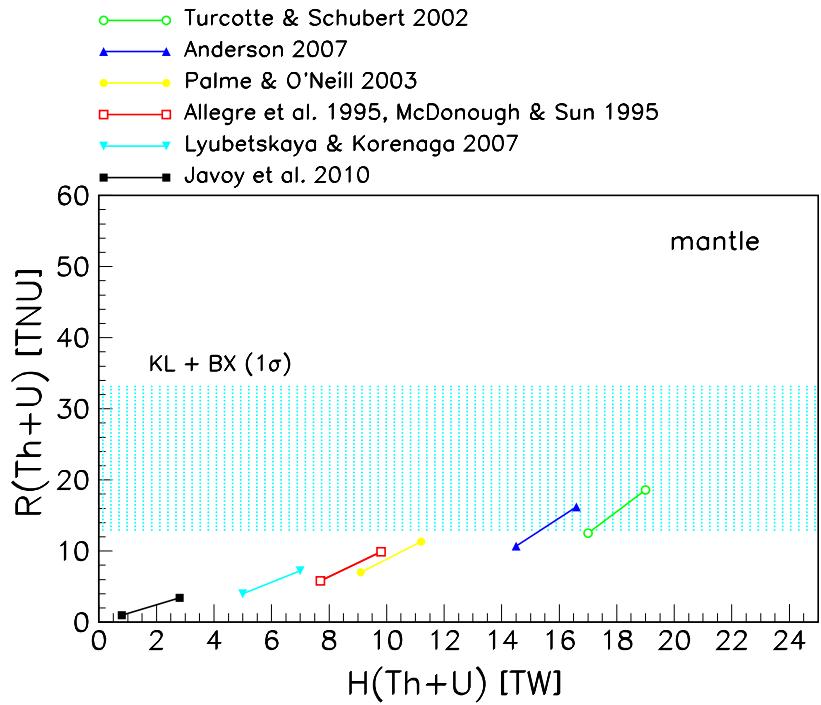}
\caption{Comparison of KL+BX constraints ($1\sigma$ horizontal band) and model predictions (slanted lines) in the plane charted by the 
Th+U geoneutrino rate and radiogenic heat for the mantle.
\label{fig5}}
\end{figure}
%%%%%%%%%%%%%%%%%%%%%%%%%%%%%%%%%%%%%%%%%%%%%%%%%%%%%%%%%%%%%%%%%%%%%%

Signal-heat correlation plots as in Fig.~5 have been suggested earlier \cite{Fi07} as a way to constrain  
the radiogenic heat, whose preferred range should be determined by 
the intersection of the slanted ``theoretical'' band (i.e., the envelope of all models) 
with the horizontal ``experimental'' band.
At present, given the large uncertainties affecting the bands 
in Fig.~5, we refrain from deriving a preferred range for the mantle heat---a task which is
left to future studies involving more accurate data and, possibly, more extensive model surveys. 
We just observe that Fig.~5 suggests a lower bound for the Th+U mantle heat, which
appears to exceed $\sim 13$~TW at $1\sigma$;
then, adding no less than $\sim 6$~TW from Th+U crustal heat (see Sec.~III~A), we infer a tentative
lower bound of $\sim 19$~TW (at $1\sigma$) for the total Th+U radiogenic heat in the Earth, compatible
with the estimate $\sim 20\pm 9$~TW of \cite{KL11}.

\section{Summary and conclusions}

In this work, we have analyzed in detail the experimental 
total rates of Th and U geoneutrino events in KamLAND (KL) and Borexino (BX), 
including neutrino oscillation effects. 
We have calculated the crustal flux at the two detector sites, 
using updated information about the global and local Th and U distribution.
After subtraction of the estimated crustal component from the total fluxes,
we find hints for residual mantle components at $\gtrsim 1.5\sigma$ in both
KL and BX. In the KL+BX combination, the statistical significance
of the mantle signal reaches the  $2.3\sigma$~level. In particular, for typical
Th/U mantle ratios, we estimate a total mantle rate of $R(\mathrm{Th}+\mathrm{U})\simeq 23\pm 10$~TNU
(including oscillation effects). The $\pm 10$~TNU error is comparable to the spread
of rate predictions derived from various published models of the mantle. 
Among these, a preference is
found for models with relatively high radiogenic contents (corresponding to present mantle
Th+U heat $\gtrsim 13$~TW at $\sim 1\sigma$). However, no model
can be excluded at $\gtrsim 2\sigma$ level yet.

The accuracy of the results can be improved in part by further KL and BX data, 
and especially by prospective data from future experiments at different sites.  
If Th and U rates from several different detectors were available,
one could estimate and subtract the crust components in all of them,
in order to infer the corresponding mantle components. Should all mantle rates  
be the same within errors, the hypothesis of a common, isotropic mantle geoneutrino
flux would be corroborated, and its value could be obtained by
combining the results from all the experiments. Conversely, one 
should consider the possibility of an anisotropic mantle flux, or of 
incorrect estimates of (some of) the subtracted crustal fluxes. 
Mantle-dominated measurements 
at oceanic sites would provide crucial tests of the various options
and additional constraints for mantle models. 
In any case, we shall learn a lot more about the Earth mantle from future
geoneutrino data, supplemented by detailed descriptions of the crust
at global and local scales, in a truly interdisciplinary approach.

\acknowledgments

This work is partly supported by the Italian
Ministero dell'Istruzione, dell'Universit\`a e della Ricerca (MIUR) through the project
Progetto di Rilevante Interesse Nazionale (PRIN) ``Fisica Astroparticellare: Neutrino ed
Universo Primordiale,'' and partly by the 
 Istituto Nazionale di Fisica Nucleare (INFN) through the  research initiative    
 ``Fisica Astroparticellare FA51.'' 
  We acknowledge useful discussions with 
 E.~Bellotti, L.~Carmignani, A.~Ianni, W.F.~McDonough, and R.~Rudnick.

\appendix
\section{Central values and uncertainties of estimated crustal rates}

In this work, a careful evaluation of crustal geoneutrino rates has been carried out (Sec.~III) in order  
to infer the mantle signal by subtraction (Sec.~IV). In this Appendix, we discuss possible effects or additional
constraints that might slightly change either the central values or the uncertainties 
of our estimated crustal fluxes; we argue that such changes 
are not expected to significantly weaken the inferred mantle signal.

\subsection{Central values}  
  
In Sec.~III, crustal rates have been estimated by summing the 
LOC (local) component, based on a detailed description of the crust near the detector
site,  and the ROC (rest-of-the-crust) component, based on state-of-the-art
information about the global structure and radiogenic contents of the crust. 

Of course, the global model of the crust used in Sec.~III can be improved by adding 
new data or constraints, which might also lead to variations in the central ROC values. 
In particular, heat-flow measurements at 
local, regional, and global scales appear to provide particularly promising 
and independent constraints \cite{Mare}. In this context, we note that a recent heat-flow evaluation  
of the total radiogenic heat of the crust suggests a value 
$H(\mathrm{Th}+\mathrm{U}+\mathrm{K})\simeq 7.5\pm 0.7$~TW \cite{Heat} (as quoted in \cite{Mare}),
which is $\sim 20\%$ lower than our estimate $8.9\pm 1.2$~TW 
 in Sec.~III, although compatible within 
the quoted $1\sigma$ errors. At present, it is not obvious 
if and how the above heat-flow estimate should be added as an additional input; 
in any case, its possible inclusion would presumably lead to
an $O(10\%)$ decrease of the adopted global Th and U crust abundances,
and to an associated decrease of the estimated ROC fluxes in both KL and BX.
The mantle flux estimated by subtraction would then slightly increase, 
and the hints for a mantle signal discussed in Sec~IV would be strenghtened.
 
Another issue concerns the
approximation of averaged oscillations. While this approximation is
certainly justified for the ROC component, it may be slightly inaccurate 
for the LOC component. We have re-calculated the LOC rates 
for the unaveraged oscillation probability in Eq.~(\ref{Pee}), 
and we find the following fractional variations with respect
to the LOC rates in Table~IV:  $+1\%$ (Th) and $+3\%$ (U) for KL,
and $-7\%$ and $-6\%$ for BX. [In BX, a larger
fraction of local flux is suppressed by the first dip of the oscillation 
probability.] 
In terms of total (ROC+LOC) rates, 
all such variations are smaller than $\pm 2.5\%$
and can be ignored in practice (see also below). 
Note that, if the effects of unaveraged oscillations were
statistically relevant for LOC estimates, then the analysis of 
geoneutrino energy spectra should also account for small $L/E$ effects
(not included in the present work). 

Based on the above considerations and on our educated  guess, 
we surmise that future 
improvements in the local and global description of the crust
are unlikely to
change our estimated crustal rates by much more than $O(10\%)$, which is also 
the size of the uncertainties in Table~IV. 
Uncertainties of this (and even twice as large) size 
are negligible in the context of the present work, as we demonstrate below.

\subsection{Uncertainties}

In Sec~IV we have inferred the mantle geoneutrino rates by subtracting 
the crust component from the total experimental rates, whose errors
have been claimed to dominate the results. The irrelevance of the
theoretical crust rate errors can be easily demonstrated in the 
case of KL, where the $n\sigma$ experimental
contours in the upper panel of Fig.~2 are nearly elliptical and equally spaced 
(i.e., the correlated errors can be approximated by a bivariate gaussian).   

In order to simplify the argument and the notation, let us call $(x,\,y)$ the
total KL experimental U and Th rates 
shown in the upper panel of Fig.~2, in TNU units. The allowed contours
are well approximated by ellipses centered at  $(27.8,\,14.5)$ and with $1\sigma$ error matrix 
%....................
\begin{equation}
\bm{\sigma}^2 = \left(
\begin{array}{cc}
\sigma^2_x & \rho\,\sigma_x\sigma_y\\
\rho\sigma_x\sigma_y & \sigma^2_y
\end{array}
\right)  \ , 
\end{equation}
%...................
where $\sigma_x=18.6$, $\sigma_y=11.4$, and $\rho=-0.75$. 
If we simply subtract the 
crustal rates $(x_c,\,y_c)\simeq (20.2,\,6.3)$ in Table~IV with no errors, we get an estimated mantle
signal $(x_m,\,y_m)=(7.6,\,8.2)$, with the same error matrix as above.  The corresponding 
$\chi^2$ function \cite{PDGB} equals $2.57\simeq 1.6^2$ at the point $(0,\,0)$, namely, 
the null hypothesis of ``no mantle signal'' is rejected at $1.6\sigma$ in this gaussian approximation
for the KL experimental errors (consistently with the non-approximated value of $1.7\sigma$ reported in Sec.~IV). 

If the KL crustal rate errors in Table~III are included with a (presumably positive) correlation $\rho_c$ in a
crust error matrix $\bm{\sigma}_c$, the mantle error matrix is augmented as
$\bm{\sigma}^2_m=\bm{\sigma}^2+\bm{\sigma}^2_{c}$.
The null hypothesis is then rejected at $1.58$--$1.59\sigma$, depending on $\rho\in[0,\,1]$. 
Therefore, the inclusion of crustal rate errors 
changes the statistical significance of the mantle signal in KL by a negligible fraction of one
standard deviation ($0.02\sigma$ or less).
If we repeat the above exercise with crustal errors doubled (or tripled) for the sake of
conservativeness, then the significance of the mantle signal in KL is lowered by less than $0.08\sigma$
(or $0.18\sigma$). For BX (middle panel of Fig.~3), the effect of crustal errors can only be smaller, since 
the experimental errors are significantly larger than in KL. Therefore, we surmise
that the $2.4\sigma$ statistical significance of the mantle signal (Sec.~IV) could be
lowered by about 0.1--$0.2\sigma$ in the worst cases.  

Summarizing, variations or uncertainties of the estimated crustal rates at the level of $O(10\%)$ 
can be practically ignored in the context of this work, whose results are dominated by the---much larger---experimental 
errors on geoneutrino event rates.  
In the future, when theoretical and experimental errors
will be comparable, a proper convolution of their distributions will be required in order
to obtain reliable error estimates for the inferred mantle geoneutrino rate.

\section{Side results of the KL and BX data analysis}

In this Appendix we supplement the data analysis reported in Sec.~II and in Fig.~1 with
additional  results, concerning total rates only, with no separation between mantle and crust components.
As in \cite{Prev}, we re-express the experimental constraints in terms of total Th+U 
rates and Th/U abundance ratios, instead of separate Th and U geoneutrino rates.  
This alternative formulation is particularly useful to evaluate the effects of assuming, in first approximation, 
the same average Th/U ratio in both KL and BX  \cite{Prev}.  The results are shown in the various panels of Fig.~5.
 
Figure~6(a) shows the best fits and $1\sigma$ contours in the plane charted by the total rate 
$R(\mathrm{Th}+\mathrm{U})$ and by the average Th/U abundance ratio. 
In the upper panel, these observables are left free for both KL and BX. Both experiments
provide similar (although weak) upper bounds on Th/U, but only KL sets a lower limit at present,
$\mathrm{Th}/\mathrm{U} \gtrsim 1$. In the lower panel, under the assumption of
 a common Th/U ratio in KL and BX,
we obtain a best fit $\mathrm{Th}/\mathrm{U}\simeq 5$, close to the typical chondritic value 
$\mathrm{Th}/\mathrm{U}\simeq 3.9$ \cite{Rati}, although with very large uncertainties (a factor of about four).

Figure~6(b) shows the constraints on the Th/U abundance ratio in terms of standard deviations 
$N_\sigma$ (the total rate being marginalized away). As in Fig.~6(a), the upper 
panel refers to the unconstrained fit, while the lower panel refers to the case with the same 
Th/U in KL and BX. It can be seen that, in combination, KL+BX  can set for the first time a $2\sigma$ upper bound
on Th/U, although it is still very weak ($\mathrm{Th}/\mathrm{U}\lesssim 10^2$). In this context, significant progress will require
geoneutrino energy spectra with high statistics and low systematics, in order to better discriminate the different Th and U components
and to constrain their ratio with higher accuracy.

Finally, Fig.~6(c) shows the constraints on the total rate $R(\mathrm{Th}+\mathrm{U})$ in terms of standard deviations 
$N_\sigma$ (the Th/U ratio being marginalized away). The upper panel coincides with Fig.~1. The
lower panel is rather similar, implying that current constraints
on the total Th+U rate are not very sensitive to specific assumptions on the Th/U ratio.

%%%%%%%%%%%%%%%% FIGURE 6 %%%%%%%%%%%%%%%%%%%%%%%%%%%%%%%%
\begin{figure}[b]
\begin{center}
\hspace*{-1.5mm}
\subfigure[\ Best fits and $1\sigma$ contours in the plane 
of the Th+U rate versus the Th/U ratio.]{
\includegraphics[scale=0.39]{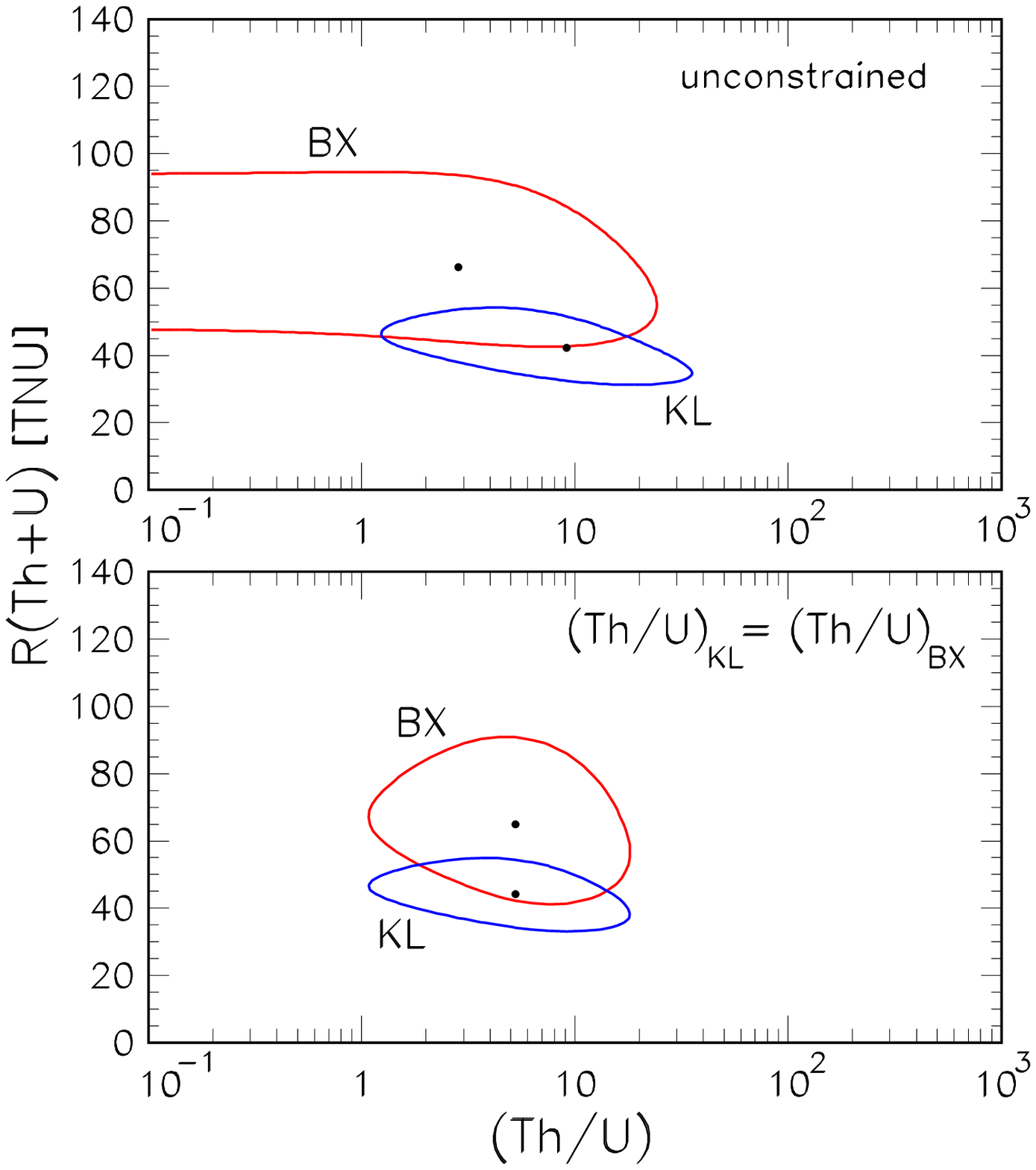}}\hspace*{1.5mm}
\subfigure[\ Constraints on the Th/U ratio in terms of standard deviations $N_\sigma$ from the best fit.]{
\includegraphics[scale=0.39]{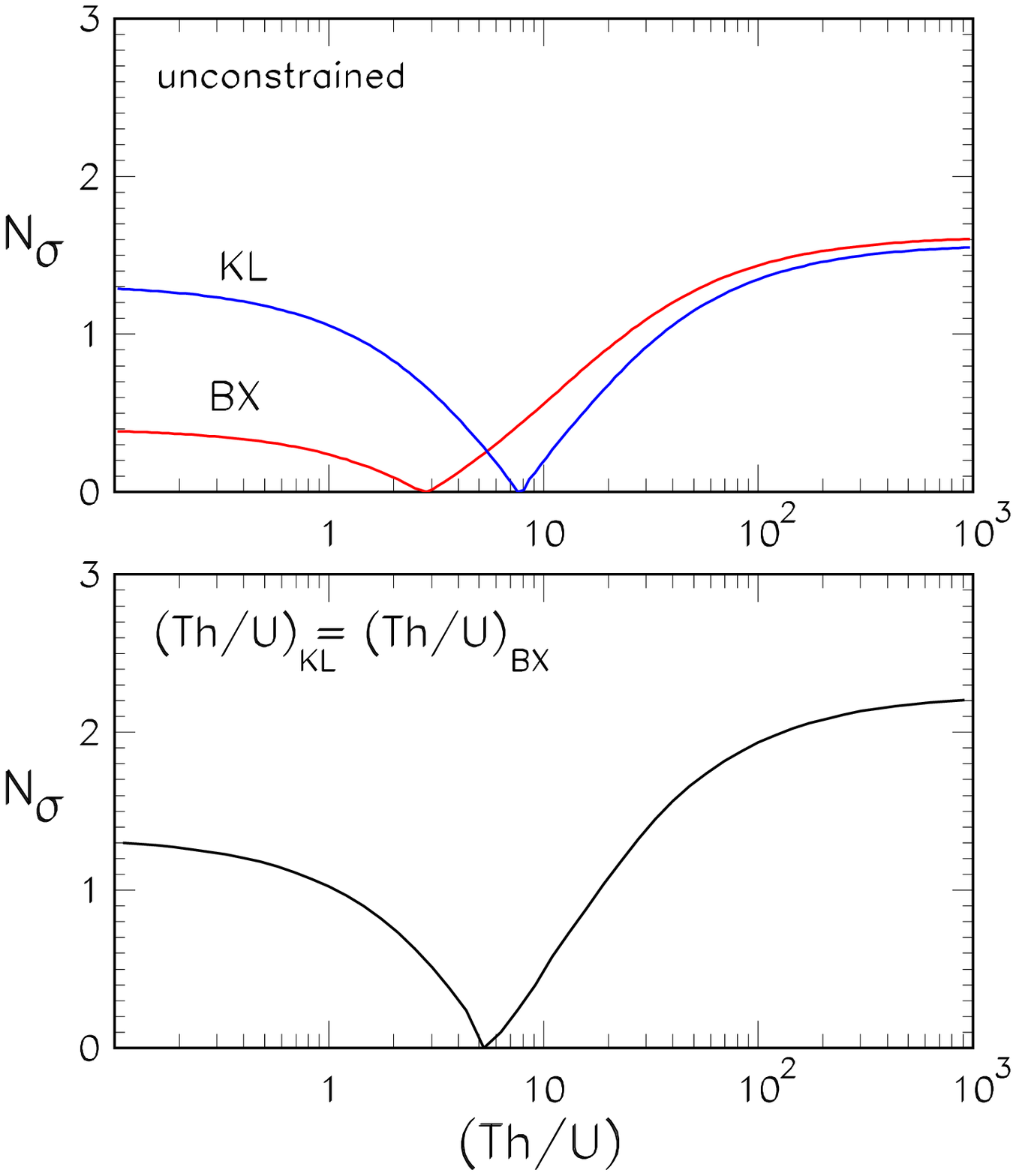}}\hspace*{1.5mm}
\subfigure[\ Constraints on the Th+U rate in terms of standard deviations $N_\sigma$ from the best fit.]{
\includegraphics[scale=0.39]{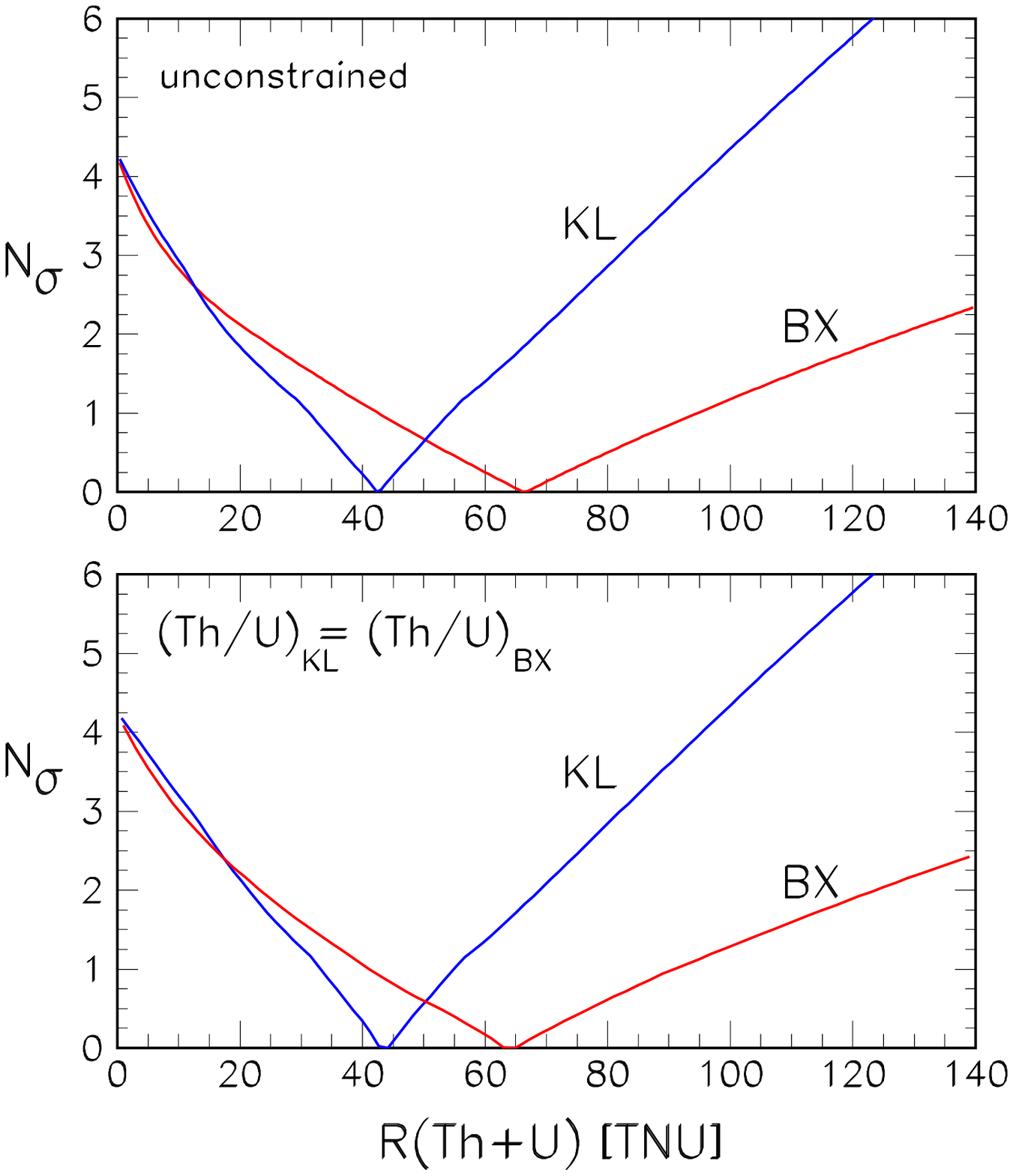}}
\caption{\label{fig6} 
Analysis of KL and BX data in terms of the total Th+U event rate and
of the average Th/U abundance ratio. In all the subfigures, the lower (upper) panel
does (not) include the assumption of equal Th/U in KL and BX.}
\end{center}
\end{figure}
%%%%%%%%%%%%%%%%%%%%%%%%%%%%%%%%%%%%%%%%%%%%%%%%%%%%%%%%%

\end{document}